# TRANSITION TO TURBULENCE IN STRONGLY HEATED VERTICAL NATURAL CONVECTION BOUNDARY LAYERS

**Thierry DE LAROCHELAMBERT**
Laboratoire G.R.E., Université de Haute-Alsace
25, rue de Chemnitz - 68200 MULHOUSE Cedex - France
e-mail: delaroche@uha.fr

**ABSTRACT**

*The mechanisms governing the transition to turbulence in natural convection boundary layers along strongly heated vertical walls remain neither very clear nor well understood, because of the lack of experiments and the difficulties of physical modelling. Our measurements bring experimental data focusing on this transition, in quiescent air along radiating and conducting plates, in the whole range of 2000 to 8000 $W.m^{-2}$ heating rate.*

*The analysis of the time series obtained by sliding window cross-correlation thermoanemometry leads us to point out coherent turbulent structures on short heights throughout the thin boundary layer, which seem to be governed by heat transfer and time microscales of turbulence through the inner sublayer. Physical interpretations are given to relate the observed heat transfer correlation and these turbulence transition structures along with radiation and conduction.*

## 1. INTRODUCTION

Fully turbulent natural convection boundary layers have been investigated at low temperatures by many experimenters, so that mean and turbulent structures are now well known, although some difficulties remain in measurement techniques, leading to scattered results in velocity and turbulent tensors profiles [1][2][3].

However, experimental investigations are somewhat scarce in high temperature or high heat flux natural boundary layers because these difficulties become numerous and particularly important, as the higher the heat flux the closer the coupling between density and velocity fluctuations, and the higher the fluctuations of refraction index [4][5][6]. Moreover, radiative and conducting effects have a greater influence on the convective heat transfer, and the way they change the intimate structures of the natural convection boundary layer during the transition to turbulence is far from obvious nor even well known.

## 2. EXPERIMENTAL DEVICE

Our experiments have tried to bring more information about such mechanisms in the transition to turbulence in natural convection boundary layers along strongly heated, vertical flat plates in quiescent air. Special attention has been paid to the measurement technique of temperature and velocity, in order to obtain reliable results despite the steep temperature and velocity gradients near the heated plates. A specific probe was built using a pair of microthermocouples (K-type, dia. 25µm), and an improved FFT-based SWICTA algorithm (sliding window cross-correlation thermoanemometry) was employed [7] *allowing simultaneous temperature and velocity measurements even in strongly fluctuating and inverting hot flows*.

Thick black steel plates (0.5m × 0.5m × 4mm) and 4000 to 8000 $W.m^{-2}$ heating rates $q_0$ were used in order to observe the coupling of radiation, conduction and convection in the heat transfer correlation, and its effect on the development and the characteristics of the turbulent natural convection boundary layer. It is to be noted that these precise heat fluxes are uniformly applied to the insulated rear face of the plates by means of thin electric ribbons in order to obtain surface temperatures around 200-350°C similar to those of the external walls of furnaces, along the free surface (*fig*. 1). More details about the experimental device, the thermoanemometric probe and the SWICTA algorithm are given in [7].

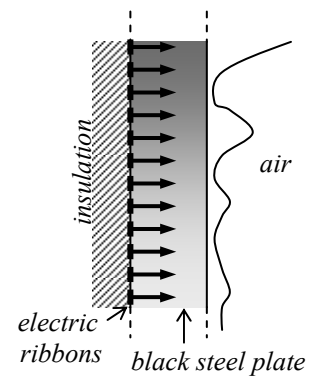

Fig. 1

Each collection of experimental measurements consisted in automatic records of quite long time series of temperatures at each point in the whole boundary layer in stationary regime along the heated plates; at some points in the ambient air of the laboratory room (*H*=5m × *L*=10m × *l*=10m) and along the plate surface. Total emissivity profiles along the plates were measured with an infrared pyrometer (0.7 < $\varepsilon_w$ < 0.97). Earlier FTIR spectrograms have proved the diffuse-grey property of their surface [8].

The measurement precision was estimated at 1K for temperatures, 1% for surface emissivities and radiative heat fluxes, 3% for convective heat fluxes and 5% for air velocities. The maximum vertical air stratification coefficient was 0.54 $K.m^{-1}$.

## 3. HEAT TRANSFER CORRELATION DURING TRANSITION

Confirmation and extension of previous results [5] about the beginning of the transition to turbulence were obtained, characterized by a rapid decrease of the critical Grashof number down to $2.10^7$ as the heating rate is increased up to 8000 W.m$^{-2}$, so that the transition to turbulence occurs at 2000 W.m$^{-2}$ at the upper part of the plate ($x/H = 0.65$), extends to the whole plate at 4000 W.m$^{-2}$ and leads to almost fully developed turbulence at the top of the plate at 8000 W.m$^{-2}$ (*fig.* 2).

The influence of radiation and conduction on the heat transfer balance was numerically and experimentally analysed in a previous work [9], but it was not clear if it concerns a part or the whole thickness of the boundary layer or if it simply lies within the gross balance of local heat fluxes along the emitting surface. The experimental heat transfer correlations $Nu_x(Gr_x)$ that had been obtained in [9] showed a characteristic behaviour, progressively departing from the laminar correlation at 2000 W.m$^{-2}$ and becoming quite turbulent at 8000 W.m$^{-2}$ after a well-defined transition correlation at 4000 W.m$^{-2}$. These results invited us to conclude that radiation leads to a weaker Nusselt number for a higher heating rate, while conduction involves heat flux diffusion in the plate that smoothes the surface temperature profiles of the plate, both phenomena promoting the earlier thermal destabilization of the boundary layer. On the other hand, the inner sublayer temperature profiles (*fig.* 4) present a small and rapid decrease very near the plate followed by a characteristic flattening that may be due to radiation absorption effects in air, thus enforcing this turbulence promotion.

However, if *the role of turbulence microscales* in the mechanism of heat exchange in the onset of and the transition to turbulence is taken into account [10], it must admitted that this mechanism prevails all along this transition and must be reflected in those heat transfer correlations. Thus, analysing turbulent heat transfers microscales in vertical natural convection boundary layers, we consider in the present report that the local Grashof number must be replaced by the non-dimensional group $Ra_x Pr_x/(1+Pr_x)$ *based on the real thermophysical properties at the film temperature $T_f = (T_w+T_0)/2$, and that the local Nusselt number $Nu_x$ must be calculated at the plate temperature $T_w$ in the very thermo-conductive sublayer*. Therefore, *we obtain the following well-fitted and single correlation (R ~ 0.995)*

$$Nu_x = 0{,}194 \left( \frac{Ra_x Pr_x}{1 + Pr_x} \right)^{0{,}282}.$$

This applies to the bulk transition process at 4000 W.m$^{-2}$ and 8000 W.m$^{-2}$ (fig 3), which is coherent with the structures examined below (§ 4), and accounts for radiative and conductive heat fluxes by removing them from the local heat flux in $Nu_x$. Moreover, this single correlation validates our hypothesis that *the thermal microscales of turbulence preside over the mechanisms during the whole transition to turbulence.*

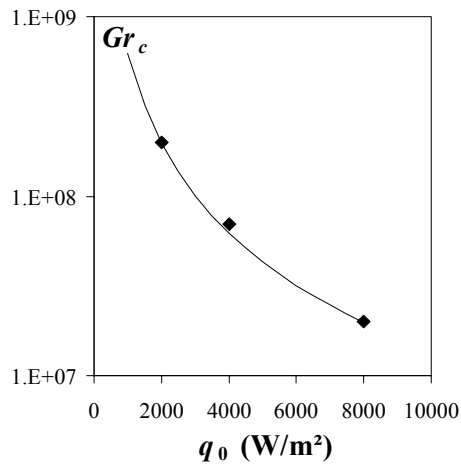 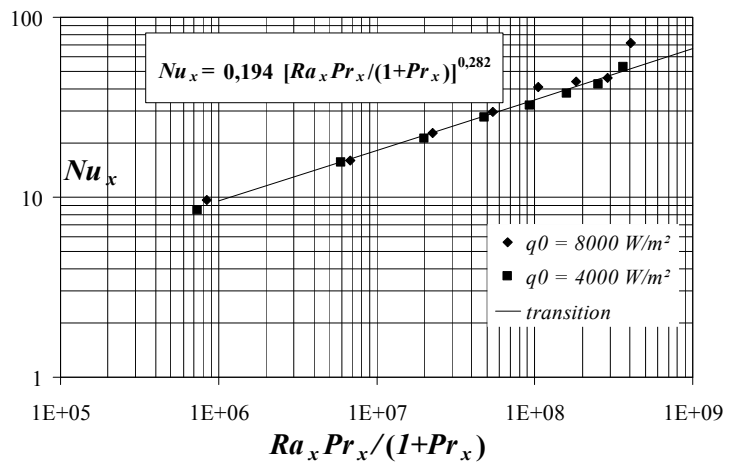

Fig. 2. *Critical Grashof number*       Fig. 3. *Heat transfer correlation during transition*

## 4. STRUCTURES AND MICROSCALES OF THE TRANSITION BOUNDARY LAYER

The progressive installation of the turbulent structures from 4000 to 8000 W.m$^{-2}$ and from the bottom to the top of the heated plate was observed through the modifications of the mean and fluctuating profiles of the air temperature, streamwise velocity and turbulent heat flux, as well as the time microscale of turbulence.

Compared to the ones that have been previously obtained at low temperature [1][2][3], these profiles present some specific shapes and distinct changes that will be examined afterwards.

### 4.1. Mean temperature and velocity profiles

Compared to the results obtained in [3], the high heat flux transition boundary layer appears less thick, growing from $\zeta = 4$ to more than 10, instead of 40 at low temperature. It is characterized by a pronounced flattening of the temperature profile in the thermo-conductive sublayer ($\zeta < 0.1$), while the streamwise velocity profile becomes progressively better-organized as the heating rate is increased from 4000 to 8000 W.m$^{-2}$, with a slight displacement of the maximum $U_m \sim 0.3\, U_b$ towards the outer sublayer from $\zeta = 1$ to 2 (*fig*.5).

After a slowly raising shape $U \sim y^3$ in the "thermo-viscous sublayer" due to the predominant role of the thermo-conductive heat flux through this sublayer and its higher viscosity [7], the velocity profile tends to follow a well-fitted power-law $U = by^{1/3} - a$ in the "buoyant sublayer" ($0.1 < \zeta < 0.3$) – according to the theoretical model in [11] –, then a positive logarithmic slope in an "inertial sublayer" and a negative one in the outer sublayer ($2 < \zeta$).

This behaviour at high heat fluxes is to be compared with the $U \sim y$ near wall profile of the fully turbulent boundary layer at low temperature [3] where the streamwise velocity reaches a maximum value around $0.34\, U_b$ located at $\zeta \sim 2$. This indicates us that during the transition at high heat flux the maximum streamwise velocity remains always slightly lower than the one at low temperature, although it reaches the same distance to the plate. The main difference between the high heat flux and the low temperature cases is the surprisingly well-organized dynamic structure of the boundary layer on short times with well-marked parts of specific shapes.

The buoyant velocity $U_b = [g\beta_0(T_w - T_0)]^{1/2}$ appears to be a well-suited reference for dimensionless velocity, which implies that *the local temperature difference* $(T_w - T_0)$ *is more physically significant than the heating flux* $q_0$ because the latter doesn't determine the local convective heat flux $q_w$ directly.

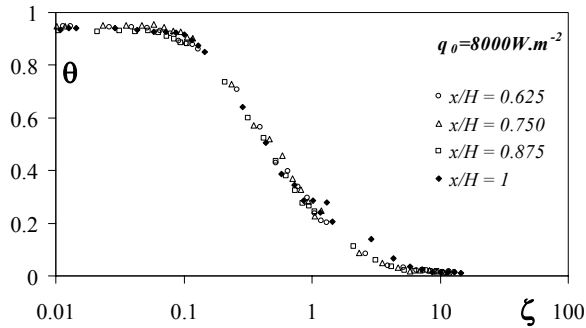
Fig. 4a: *Thermal profile at 8000 W.m$^{-2}$*

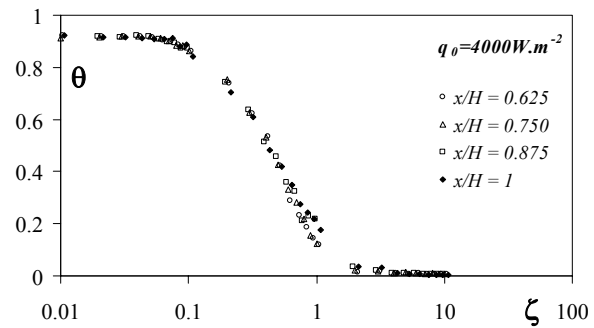
Fig. 4b: *Thermal profile at 4000 W.m$^{-2}$*

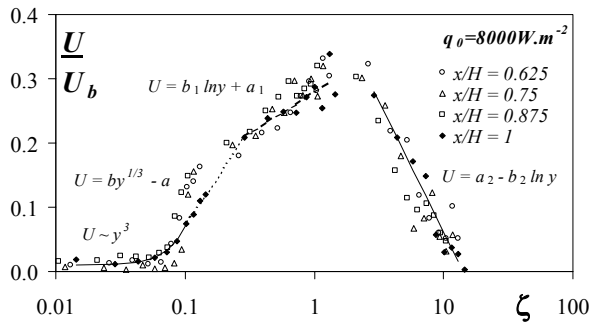
Fig. 5a: *Streamwise velocity profile at 8000 W.m$^{-2}$*

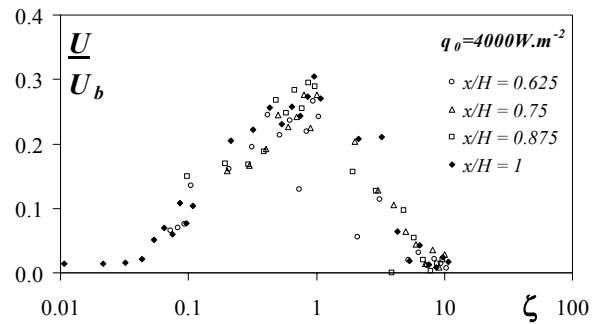
Fig. 5b: *Streamwise velocity profile at 4000 W.m$^{-2}$*

### 4.2. Intensities of temperature and velocity fluctuations

The progressive organization of the boundary layer during the transition process is more obvious if one considers the intensity profiles of the temperature and velocity fluctuations. Here again, the thickness of the fluctuating boundary layer at high heat flux during transition is smaller than the turbulent one at low temperature in [3]. If the intermittency is important enough at 4000 W.m$^{-2}$ to upset an ordered structure, higher heating rates have a real and efficient role in the production of coherent fluctuating structures.

It appears that the maximum value of the thermal turbulence intensity stabilizes around 0.12-0.13 at $\zeta \sim 0.6$-0.7 during transition (*fig*. 6) but remains relatively important in the thermo-viscous sublayer, which is a sign of a destabilization of this sublayer by the more violent eddies coming from the outer layer. This maximum is lower than the 0.17 value obtained at the same location in the fully turbulent boundary layers at low temperature in [3]; this may be due to a higher destruction rate of kinetic energy in the inner sublayer (see § 4.3 below).

This destabilization is more pronounced for the turbulence of velocity, as can be seen on *fig.* 7: the maximum intensity of the fluctuating velocity tends to shift in the outer sublayer at about $\zeta \sim 8$ around a value of 0.13 $U_b$ when $q_0$ is increased up to 8000 W.m$^{-2}$, and the approximately logarithmic slope of these fluctuations towards the plate is low, so that they remain important at the edge of the thermal sublayer. This may be due to the strong density fluctuations during transition. This explains that the velocity fluctuation profiles during transition, rather scattered at 4000 W.m$^{-2}$, remain higher than those in [3], with almost similar shapes at 8000 W.m$^{-2}$.

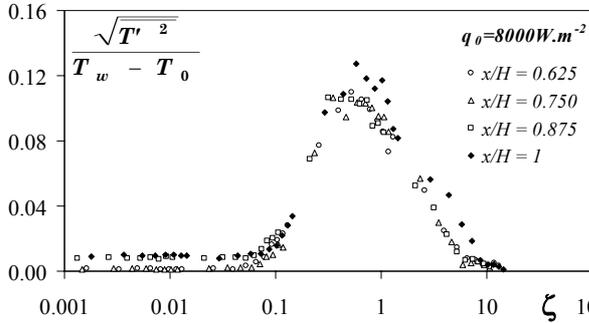
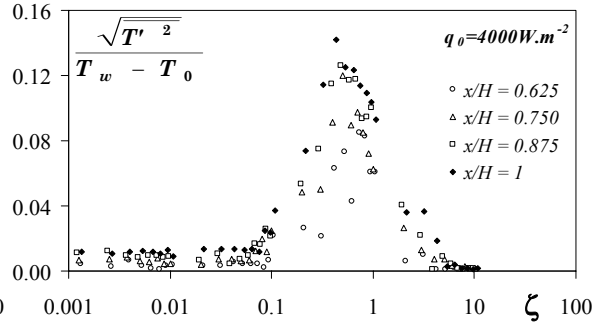

Fig. 6a: *Intensity of thermal fluctuations at 8000 W.m$^{-2}$*    Fig. 6b: *Intensity of thermal fluctuations at 4000 W.m$^{-2}$*

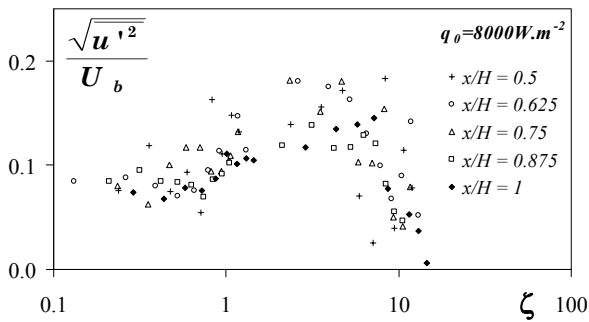
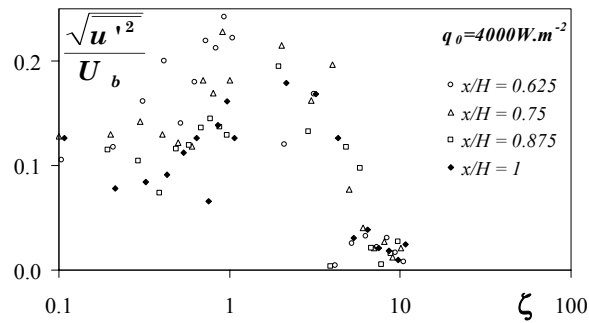

Fig. 7a: *Intensity of velocity fluctuations at 8000 W.m$^{-2}$*    Fig. 7b: *Intensity of velocity fluctuations at 4000 W.m$^{-2}$*

### 4.3. Turbulent vertical heat flux profiles

The way the convective heat flux is carried throughout the inner sublayer and structures the whole boundary layer is revealed by the profiles of streamwise turbulent heat flux. The peculiar tendency observed at 4000 W.m$^{-2}$ is fully confirmed at 8000 W.m$^{-2}$, the profiles being characterized by a significant negative value in the major part of the inner sublayer (*fig.* 8).

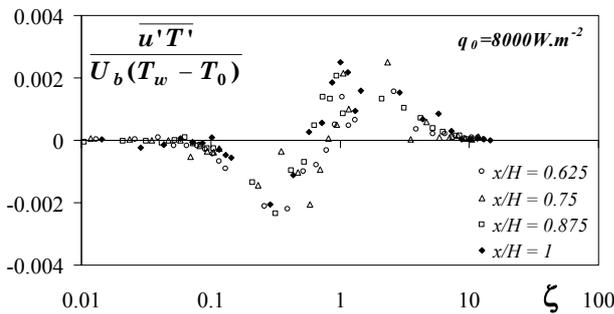
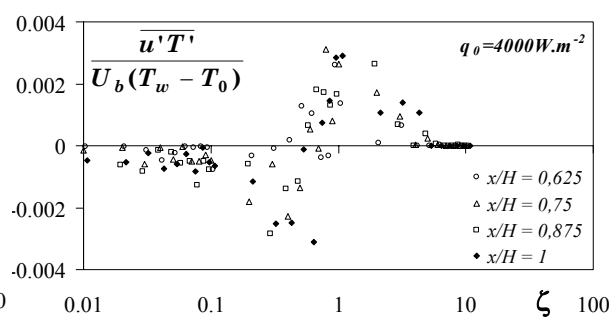

Fig. 8a: *Streamwise turb. heat transfer at 8000 W.m$^{-2}$*    Fig. 8b: *Streamwise turb. heat transfer at 4000 W.m$^{-2}$*

Although this property is quite non-existent at low [3] or high [6] temperature in the fully turbulent boundary layer, it has been numerically predicted previously [12]. It seems that experimental difficulties make these measurements particularly sensitive to a close correlation between both fluctuating thermal and dynamical signals, especially near hot walls. With our method allowing simultaneous measurement of temperature and velocity fluctuations, the SWICTA algorithm seems to give more reliable results in this very case.

This negative value of the $\overline{u'T'}$ tensor implies a positive value of the $\overline{\rho'u'}$ tensor in the inner sublayer, leading to the destruction of turbulent kinetic energy by buoyancy, the energy being transferred to the mean flow with a strong acceleration in this part of the boundary layer as observed above. This phenomenon is perhaps to be compared with the laminarization observed in combined-convection boundary layers [13].

### 4.4. Time microscale of thermal turbulence

The rate of destruction of small eddies is related to the previous mechanism and prevails in the whole part of the boundary layer where thermal energy is converted into mechanical energy by buoyancy forces. The fundamental parameter of such a mechanism is the time microscale of turbulence $\lambda_{T,t}$ which can be defined from the auto-correlation function as the de-correlation time of small thermal eddies at each point of the boundary layer

$$R_T(t) \approx 1 - \frac{t^2}{\lambda_{T,t}^2}.$$

A characteristic time scale of this destruction process has to be found by dimensional analysis on physical grounds. Considering that the principal parameters are the thermal diffusion $a$, the gravitational acceleration $g$, the thermal expansion $\beta$ and the temperature difference $\Delta T = (T_w - T_0)$, we obtain a time scale of destruction of the kinetic energy by thermal diffusion

$$\tau_T = \left(\frac{a}{(g\beta\Delta T)^2}\right)^{1/3}.$$

Thus, similarity is obtained for the $\lambda_{T,t}/\tau_T$ ($\zeta$) profiles at 4000 and 8000 W.m$^{-2}$ (*fig*. 9). It can be seen that $\lambda_{T,t} \sim 6\tau_T$ in the inner part of the boundary layer, according to the above analysis, while increasing up to 30 $\tau_T$ in the outer part where velocity decreases and coarse cooled eddies are carried out for long distances in quiescent air. Unfortunately, no other experimental data are available in the literature to our knowledge, even at low temperature.

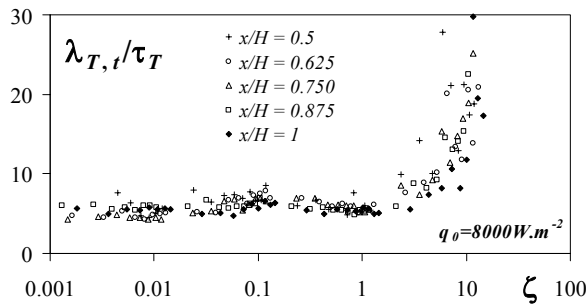
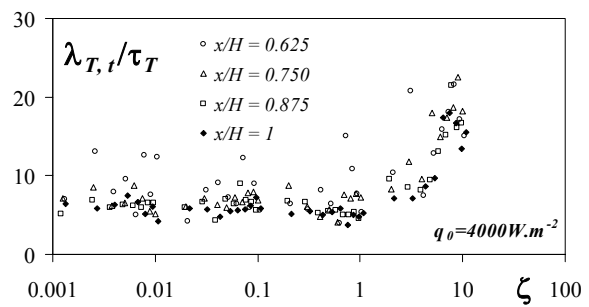

Fig. 9a: *Time microscale of thermal turbulence at 8000 W.m$^{-2}$*

Fig. 9b: *Time microscale of thermal turbulence at 4000 W.m$^{-2}$*

### 5. CONCLUSION

The behaviour of the natural convection boundary layer at high heat flux during the transition to turbulence shows characteristic tendencies and progressively coherent structures that are very quickly established on short distances, leading to relatively thin boundary layers.

The comparison with the well-known fully turbulent structures of the natural convection boundary layer at low temperature brings a better understanding of the significant modifications that are induced in the mean and fluctuating scales by strong heat fluxes.

The thermo-conductive sublayer plays a predominant role in this organization that seems responsible not only for the shape of mean velocity profiles but also for the whole heat transfer correlation during the transition. It also appears to induce an homogeneous process of thermal eddies destruction through the inner sublayer, which explains the negative turbulent heat flux as well as the lowering of the turbulence intensity of the velocity in this sublayer as the heating rate is increased.

Even if radiative heat transfers dominate when high heating rates are applied to conducting black plates and participate in the destabilization of the boundary layer as well as conductive diffusion of heat fluxes, it has been shown that the turbulent structures are self-organized and governed by homogeneous thermal turbulent microscales.

These characteristic phenomena which occur during the transition to turbulence at high heating rates lead to a unique behaviour of the heat transfer correlation if thermal turbulent microscales are taken into account.

## 6. NOMENCLATURE

| | | | |
|---|---|---|---|
| $a$ | thermal diffusivity (m$^2$.s$^{-1}$) | $x$ | vertical distance from leading edge (m) |
| $Gr_x$ | local Grashof number $g\beta_0(T_w-T_0)x^3/\nu^2$ | $y$ | horizontal distance from the plate (m) |
| $g$ | gravitational acceleration (m.s$^{-2}$) | | |
| $H$ | height of the plate (m) | | *Greek symbols* |
| $h_c$ | local convective heat transfer coefficient | $\beta$ | thermal expansion coefficient (K$^{-1}$) |
| $I_T$ | intensity of thermal fluctuations $\sqrt{\overline{T'^2}}/(T_w-T_0)$ | $\lambda_{T,t}$ | time microscale of thermal turbulence (s) |
| $I_U$ | intensity of velocity fluctuations $\sqrt{\overline{u'^2}}/U_m$ | $\lambda$ | thermal conductivity (W.m$^{-1}$.K$^{-1}$) |
| $Nu_x$ | local Nusselt number $h_c x/\lambda_w$ | $\theta$ | dimensionless temperature $(T-T_0)/(T_w-T_0)$ |
| $Pr_x$ | local Prandtl number $\nu_f/a_f$ | $\nu$ | kinematic viscosity (m$^2$.s$^{-1}$) |
| $q$ | heat flux (W.m$^{-2}$) | $\tau_T$ | thermal diffusion time microscale $[a/(g\beta\Delta T)^2]^{1/3}$ |
| $R$ | correlation function or coefficient | $\zeta$ | dimensionless distance from wall $-y\left(\dfrac{\partial\theta}{\partial y}\right)_w$ |
| $Ra$ | local Rayleigh number $g\beta_0(T_w-T_0)x^3/\nu_f a_f$ | | |
| $T$ | temperature, °C ou K | | *Subscript* |
| $T'$ | temperature fluctuation (°C or K) | $c$ | convective |
| $U$ | vertical velocity (m.s$^{-1}$) | $f$ | at film temperature $(T_w+T_0)/2$ |
| $u'$ | vertical velocity fluctuation (m.s$^{-1}$) | $w$ | relative to the wall |
| $U_b$ | buoyant velocity $[g\beta(T_w-T_0)x]^{1/2}$ | $0$ | boundary condition |